\documentclass[aps,prb,twocolumn,preprintnumbers,superscriptaddress,amsmath,amssymb]{revtex4-2}

\usepackage{graphicx}
\usepackage{amsmath}
\usepackage{wasysym}
\usepackage{float}
\usepackage{color}
\usepackage{miller} % Miller-Indizes z.B. \hkl [1 -1 0]
\usepackage[english]{babel}
\usepackage{upgreek}
\usepackage[hidelinks]{hyperref}
\usepackage[all]{hypcap}
\usepackage{xcolor}
\usepackage{comment}
\hypersetup{colorlinks, linkcolor={black}, citecolor={black}, urlcolor={blue!50!black}}

\def\comma{, }

\usepackage{mathtools}
\DeclarePairedDelimiter\abs{\lvert}{\rvert}
\begin{document}
\title{Homogeneous doping of epitaxial graphene by Pb(111) islands: A magnetotransport study}

\author{Julian Koch}\email{julian.koch@physik.tu-chemnitz.de}
\affiliation{Institut f\"ur Physik, Technische Universt\"at Chemnitz\comma Reichenhainerstr.\ 70, 09126 Chemnitz, Germany}

\author{Sergii Sologub}
\affiliation{Institut f\"ur Physik, Technische Universt\"at Chemnitz\comma Reichenhainerstr.\ 70, 09126 Chemnitz, Germany}
\affiliation{Institute of Physics, National Academy of Sciences of Ukraine, Nauki avenue 46, 03028 Kyiv, Ukraine}

\author{Dorothee Sylvia Boesler}
\affiliation{Institut f\"ur Physik, Technische Universt\"at Chemnitz\comma Reichenhainerstr.\ 70, 09126 Chemnitz, Germany}

\author{Chitran Ghosal}
\affiliation{Institut f\"ur Physik, Technische Universt\"at Chemnitz\comma Reichenhainerstr.\ 70, 09126 Chemnitz, Germany}

\author{Teresa Tschirner}
\affiliation{Physikalisch-Technische Bundesanstalt, Bundesallee 100, 38116 Braunschweig, Germany}

\author{Klaus Pierz}
\affiliation{Physikalisch-Technische Bundesanstalt, Bundesallee 100, 38116 Braunschweig, Germany}

\author{Hans Werner Schumacher}
\affiliation{Physikalisch-Technische Bundesanstalt, Bundesallee 100, 38116 Braunschweig, Germany}

\author{Christoph Tegenkamp}\email{christoph.tegenkamp@physik.tu-chemnitz.de}
\affiliation{Institut f\"ur Physik, Technische Universt\"at Chemnitz\comma Reichenhainerstr.\ 70, 09126 Chemnitz, Germany}

\date{\today}

\begin{abstract}

Proximity coupling is an effective approach for the functionalization of graphene. However, graphene's inertness inhibits the adsorption of closed films, thus favoring island growth, whose inhomogeneity might be reflected in the induced properties. In order to study the homogeneity of the doping profile induced by an inhomogeneous coverage and the spin orbit coupling (SOC) induced in graphene, we deposited Pb(111) islands with an average coverage of up to 30\;ML on monolayer graphene (MLG) on SiC(0001) at room temperature (RT). 
We investigated the transport properties and the structure using magnetotransport, and scanning tunneling microscopy and low energy electron deflection, respectively. 
%A transition from the growth of small hexagonal islands with a width of roughly 8\;nm to the growth of larger islands with a greater variance in size was observed at a Pb coverage of about 5\;ML, which also roughly corresponds to the percolation threshold. 
The Pb(111) islands act as donors, increasing the electron concentration of graphene by about $5\times10^{11}\;\text{ML}^{-1}\text{cm}^{-2}$. The doping was found to be homogeneous, in stark contrast to our previous results for Bi islands on MLG. Upon percolation of the Pb layer at around 5\;ML, hole transport through the Pb islands has to be taken into account in order to describe the transport data. The Pb(111) islands do not induce any Rashba SOC, contrary to theoretical predictions for an interface between Pb(111) and graphene. Moreover, they seem to screen the defects in the graphene, resulting in a reduction of the intervalley scattering rate up to 5\;ML. 
\end{abstract}
%\pacs{73.50.Jt, 68.43.-h, 75.30.Hx, 73.50.-h, 73.61.-r}
\maketitle
%################################################################################################################

\begin{figure*}[tb]
	\begin{center}
		\includegraphics[width=1\textwidth]{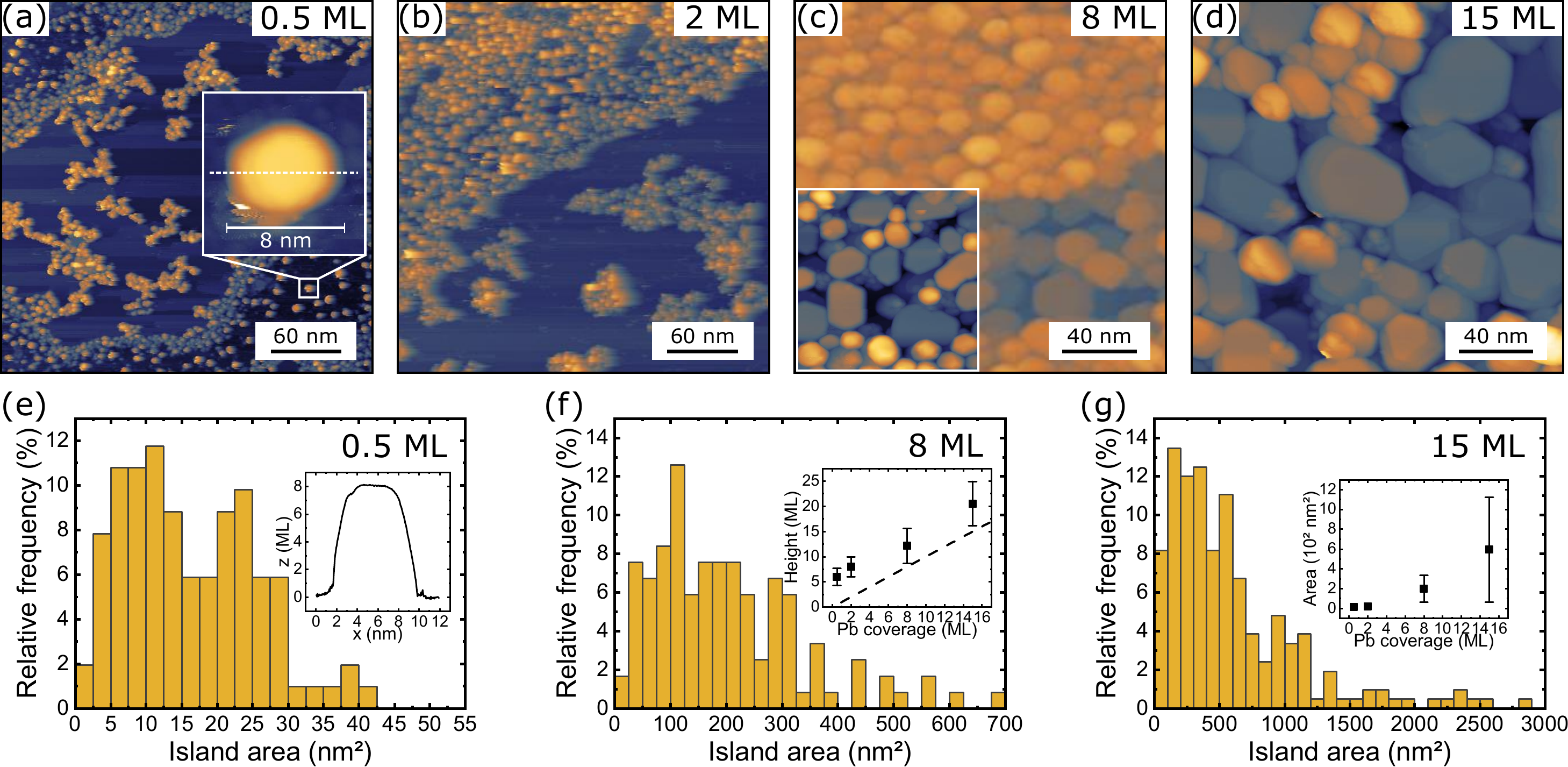}
		\caption{\label{fig:STM} 
		(a--d) STM images for average Pb coverages of 0.5, 2, 8 and 15\;ML, respectively, deposited at 300\;K. The inset in (a) shows a high resolution scan of the marked island. The inset in (c) shows a high resolution scan of that area, with the same scale as the main image. Tunneling parameters: (a) incl.\ inset 2\;V, 300\;pA; (b) 2\;V, 150\;pA; (c, d) incl.\ inset 1\;V, 110\;pA.
		(e--g) Histograms of the island area for 0.5, 8 and 15\;ML, respectively. Insets in (e--g): Height profile along the dashed line in (a), average island height and area in dependence of the total Pb coverage, respectively. The error bars represent the standard deviations. The dashed line in the inset of (f) represents the height of a closed film at the given coverage. % ($\text{Pb(111) monolayer distance} = 2.92\;\text\AA$). %2.915927
		%73-26
		}
	\end{center}
\end{figure*}

%_INTRODUCTION________________________________________________________________________________________________________________________________________
\section{Introduction}

Proximity coupling enables the tailoring of the electronic and magnetic properties of two dimensional materials. Of particular interest is the functionalization of graphene, with its unique band structure featuring Dirac cones. 
The spin lifetimes in graphene can reach values in the order of 1\;ns making it an ideal material for application in spintronics~\cite{Han2012}. However, the negligible intrinsic spin orbit coupling (SOC) in graphene, associated with a gap of only about 24\;\textmu eV~\cite{Gmitra2009}, limits the generation and manipulation of spin-polarized currents. This resulted in efforts to increase graphenes SOC by bringing it in contact with different materials such as topological insulators~\cite{Zhang2014,Jin2013,Rajput2016}, transition metal dichalcogenides~\cite{Wakamura2022,Avsar2014} and heavy metal adatoms~\cite{Weeks2011,Balakrishnan2014}, which increased the SOC gap by three orders of magnitude to values between 17 and 80\;meV. 

In this context, we recently investigated the effects of Bi(110) islands on epitaxial graphene~\cite{Koch2024}. However, the adsorbed islands resulted in a highly inhomogeneous carrier concentration. In the areas not covered by Bi, the properties of graphene were almost unaffected, except for additional scattering at the island edges, thereby limiting any potential increase of the SOC to the areas below the islands. Due to the inertness of graphene, i.e., the lack of dangling bonds, the formation of closed adsorbate layers is generally energetically unfavorable, so that island formation is the norm. With this in mind, the question arises under what circumstances an inhomogeneous coverage also causes an inhomogeneity of the properties of graphene, which may limit the effectiveness of proximity coupling, but can also be advantageous for certain applications. 

The focus of this work is on the effects of Pb islands on the electronic transport properties of monolayer graphene (MLG) on SiC(0001). Pb is located in the periodic table right next to Bi and thus has a similar mass. However, its Fermi wavelength $\lambda_F\approx3.7\;\text\AA$~\cite{Hinch1989} is two orders of magnitude smaller than that of Bi ($\sim30\;\text{nm}$~\cite{Cohen1961,Ogrin1966,Liu1995}) and, in contrast to Bi, there is only weak backscattering of the graphene electrons at the edges of Pb islands~\cite{Cherkez2018}. Therefore, we expect the effect of Pb islands on the transport properties of graphene to be significantly different from that of Bi islands. 
Deposition of the Pb or annealing at room temperature (RT) was found by Liu \textit{et al.}\ to result in (111)-faceted islands~\cite{Liu2013}. These islands are quasi-free-standing and are most stable if the number of atomic layers is even~\cite{Dil2007,Liu2013}. 
Intercalated Pb islands between graphene and Ir(111) have been shown to induce strong SOC in graphene~\cite{Calleja2014}. Theoretical calculations predict that adatoms in hollow positions increase the intrinsic SOC, while adatoms in top positions, random distributions of adatoms and structures commensurate with large graphene unit cells induce Rashba-type SOC~\cite{Brey2015}. Since the preferred adsorption site of Pb is the top position~\cite{Hupalo2011}, both individual Pb adatoms and Pb(111) islands are expected to induce Rashba type SOC. 

We employ magnetotransport measurements in order to derive the characteristic scattering times from the weak (anti)localization (WL/WAL) behavior, as well as the carrier concentration and the mobility of charge carriers for average Pb coverages of up to 30 monolayer (ML) deposited at RT. This analysis is supported by structural investigations by low energy electron diffraction (LEED) and scanning tunneling microscopy (STM). We will show that adsorbed Pb islands, in contrast to Bi islands, result in a homogeneous carrier concentration and mobility in the graphene layer. Furthermore, the Pb(111) islands do not induce Rashba-type SOC in graphene, contrary to the initial expectations.

\begin{comment}
+MLG=monolayer graphene
+LEED
+STM
+ML=monolayer
+WL=weak localization +WAL
B=magnetic field
+SOC=spin orbit coupling
\end{comment}

%_EXPERIMENTAL SETUP___________________________________________________________________________________________________________________________________
\section{Experimental methods}

All measurements and the deposition of the Pb islands were performed under ultra high vacuum (UHV) at a base pressure of at least $1\times 10^{-10}$\;mbar. 
Epitaxial monolayer graphene on SiC, grown \textit{ex situ} by polymer assisted Si sublimation from the SiC crystal in an Ar atmosphere inside a tube furnace, was used as a substrate~\cite{Kruskopf2016}. For the magnetotransport and LEED measurements $10\times10\times0.5$\;mm$^3$ 6H-SiC(0001) (semi-insulating, $\rho=2.6\times10^{11}\;\Omega\,\text{cm}$) wafer pieces were used, while the STM measurements were performed in a separate UHV chamber using a $10\times5\times0.5$\;mm$^3$ 4H-SiC(0001) (nitrogen doped, n-type, $\rho=0.0178\;\Omega\,\text{cm}$) wafer piece. The substrates were degassed inside the UHV chambers at $500^\circ$C in order to remove oxygen and other contaminants. 
Pb islands were grown \textit{in situ} at a substrate temperature of 300\;K by Pb evaporation from a Knudsen cell at a deposition rate of 0.1\;ML/min. The total coverage was controlled using a quartz crystal microbalance. 

Magnetotransport measurements were performed at a sample temperature of 12\;K and using a 4\;T superconducting split coil magnet. The magnetoresistivity and Hall resistivity were measured using an eight-probe setup based on the van der Pauw geometry, the details of which are described elsewhere~\cite{Koch2024}. 
LEED measurements were performed using a spot profile analyzing LEED (SPA-LEED). For all LEED images shown, the distortion was corrected using LEEDCal~\cite{LEEDCal,Sojka2013}. 
The STM measurements were done at 78\;K in a low temperature STM (LT-STM) using a tungsten tip. The tip was grounded during the measurements, i.e., a positive bias voltage probes the unoccupied states.

%_RESULTS AND DISCUSSION_______________________________________________________________________________________________________________________________
\section{Results and discussion}

\begin{figure}[tb]
	\begin{center}
		\includegraphics[width=1\columnwidth]{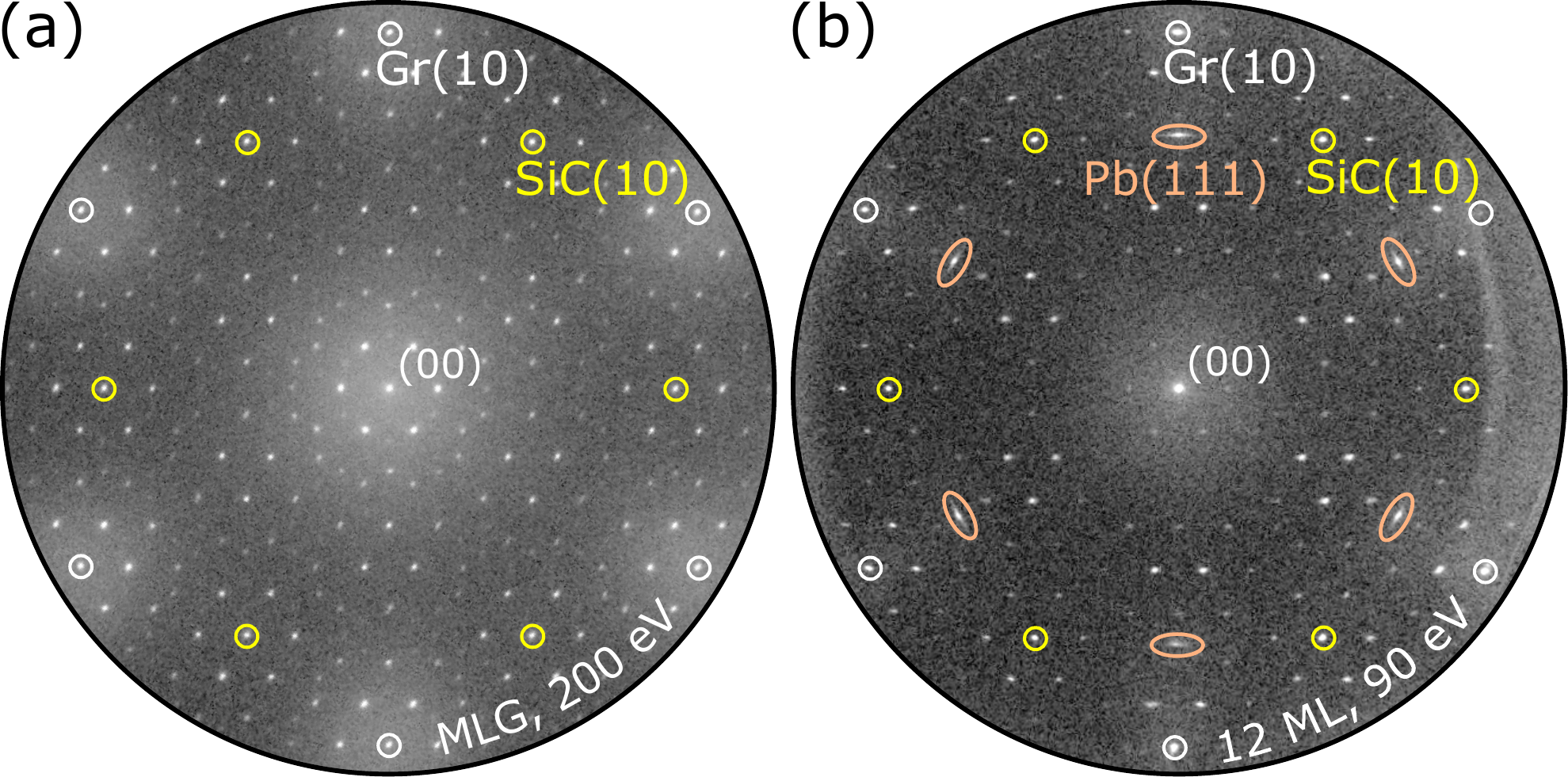}
		\caption{\label{fig:LEED}
		  (a) LEED image of the clean MLG/SiC(0001) substrate recorded at an electron energy of 200\;eV shown for reference. 
		  (b) LEED image for an average Pb coverage of 12\;ML deposited at 300\;K recorded at an electron energy of 90\;eV and a sample temperature of 230\;K. 
		%The image distortion was corrected using LEEDCal~\cite{LEEDCal,Sojka2013}.
		% G1745 (12ML) / G1774 (MLG)
		}
	\end{center}
\end{figure}

\subsection{Structural investigation}

Figures~\ref{fig:STM}(a)--(d) show STM images for average Pb coverages of 0.5, 2, 8 and 15\;ML, respectively. With increasing coverage both the covered substrate area and the average island size increase. The relative area covered by Pb is roughly 10\%, 25\%, 67\% and 75\% at 0.5, 2, 8 and 15\;ML, respectively. The percolation threshold is between 2 and 5\;ML. The average island area, as determined from representive STM images, increases from approximately 16\;nm$^2$ at 0.5\;ML to 200\;nm$^2$ at 8\;ML to 594\;nm$^2$ at 15\;ML. Detailed distributions of the island area are shown in the histograms in Figs.~\ref{fig:STM}(e)--(g). 
The orientation of the edges of the larger islands in Figs.~\ref{fig:STM}(c) and (d) indicate a crystalline structure with a hexagonal symmetry, which is confirmed by LEED to be due to a Pb(111) surface orientation as shown in Fig.~\ref{fig:LEED}(b). According to the LEED data, the lattice constant of the adsorbed Pb(111) islands is 3.5\;{\AA}, which coincides with the respective bulk value. 
At 0.5 and 2\;ML no adsorbate spots were observed in LEED. However, in a high resolution STM image (see inset of Fig.~\ref{fig:STM}(a)) the islands nevertheless appear to have a hexagonal symmetry. The height profile shown in the inset of Fig.~\ref{fig:STM}(e) reveals that the small islands have a flat top. This suggests that Pb(111) islands are present even at the lowest coverages, and that the absence of the corresponding LEED spots is simply due to their limited size. 

The average island height shown in the inset of Fig.~\ref{fig:STM}(f) approaches the height expected for layer-by-layer growth with increasing coverage, but always remains slightly larger even at 15\;ML, i.e., the Pb layer becomes never fully closed. 
The average island area increases superlinear, as shown in the inset of Fig.~\ref{fig:STM}(g). At low coverages the islands are hexagonal with only small variations in size, possibly due to strain effects. Moreover, the average island area remains almost unchanged with increasing coverage. Above a critical coverage between 2 and 8\;ML the average island area starts increasing rapidly with increasing coverage, indicating that the coalescence of several smaller islands into single larger islands becomes energetically favorable. 
We did not observe a clear preference for even-numbered monolayer heights, as reported for Pb(111) islands on graphite~\cite{Dil2007} and on graphene/Ru(0001)~\cite{Liu2013}.

\subsection{Magnetotransport}

\begin{figure}[tb]
	\begin{center}
		\includegraphics[width=.85\columnwidth]{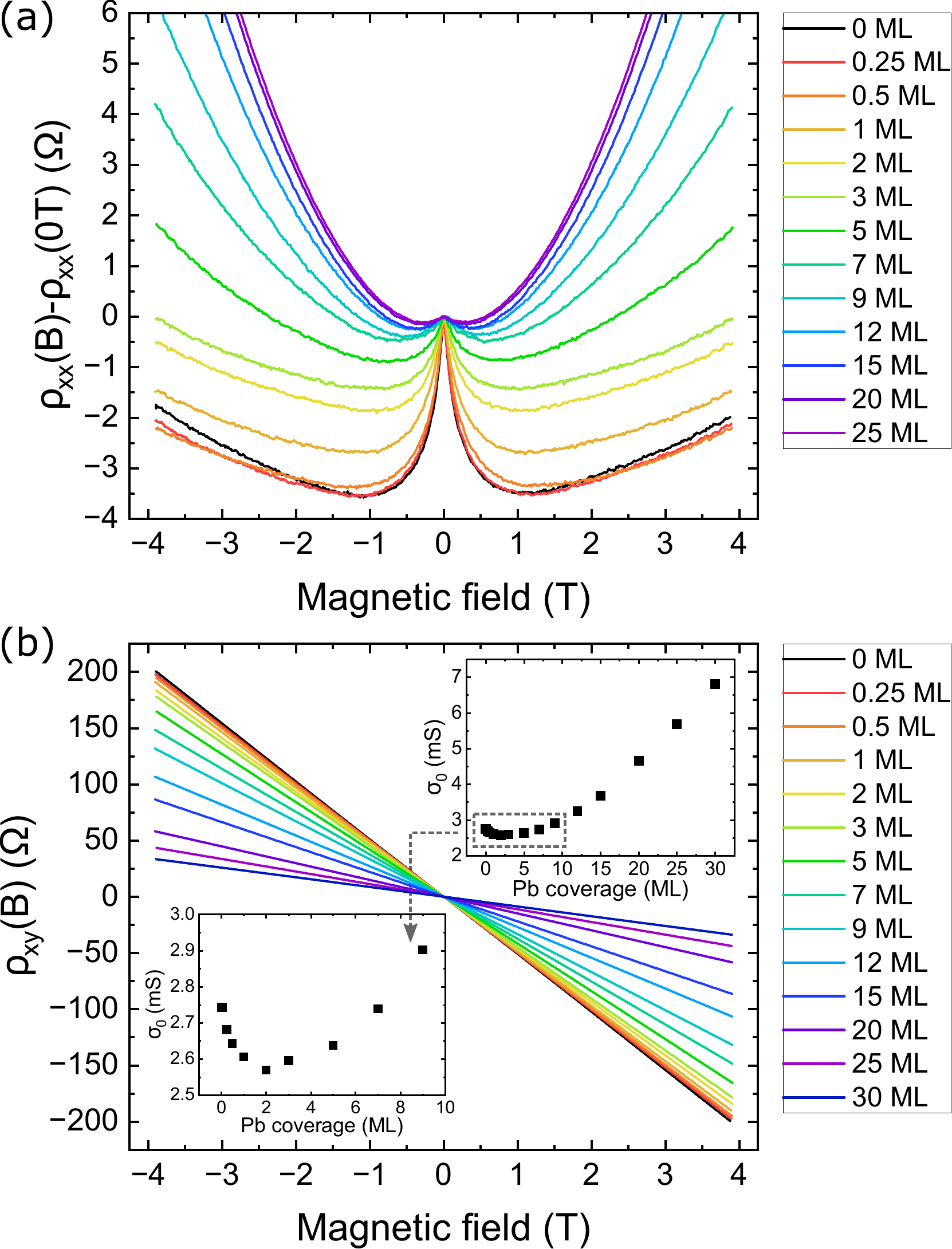}
		\caption{\label{fig:Magneto_overview} 
		(a) Magnetoresistivity and (b) Hall resistivity for average Pb coverages ranging from 0 to 30\;ML at 12\;K. For better visibility the 30\;ML curve is not shown in (a), since it overlaps with the 20 and 25\;ML curves. The top right inset in (b) shows the dependence of the conductivity at zero magnetic field $\sigma_0$ on the Pb coverage. The bottom left inset is a magnification of the same data at low coverages to highlight the minimum. 
		%G1745
		}
	\end{center}
\end{figure}

\begin{figure}[tb]
	\begin{center}
		\includegraphics[width=1\columnwidth]{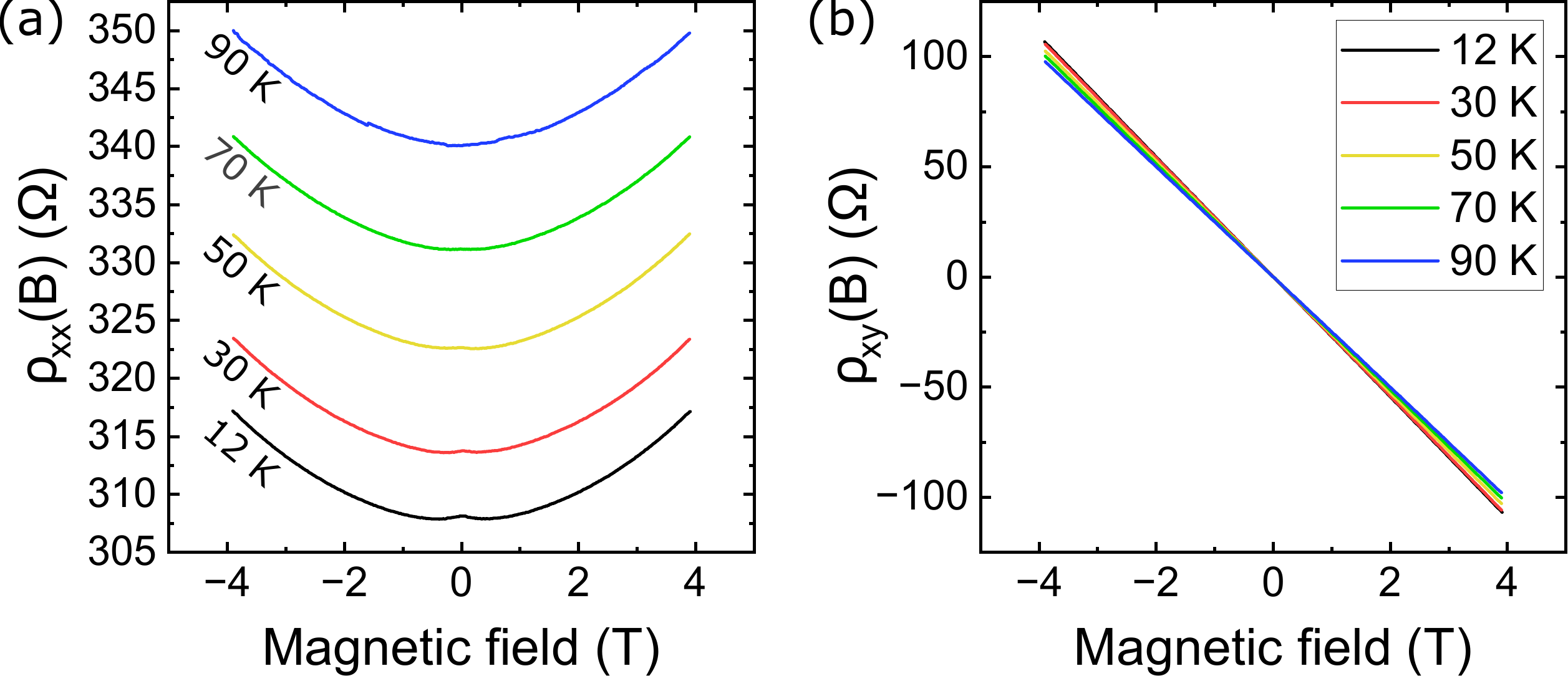}
		\caption{\label{fig:Magneto_T-dependence} 
		(a) Magnetoresistivity and (b) Hall resistivity for a Pb coverage of 12\;ML at different temperatures ranging from 12 to 90\;K.
		%G1745
		}
	\end{center}
\end{figure}

\begin{figure}[tb]
	\begin{center}
		\includegraphics[width=1\columnwidth]{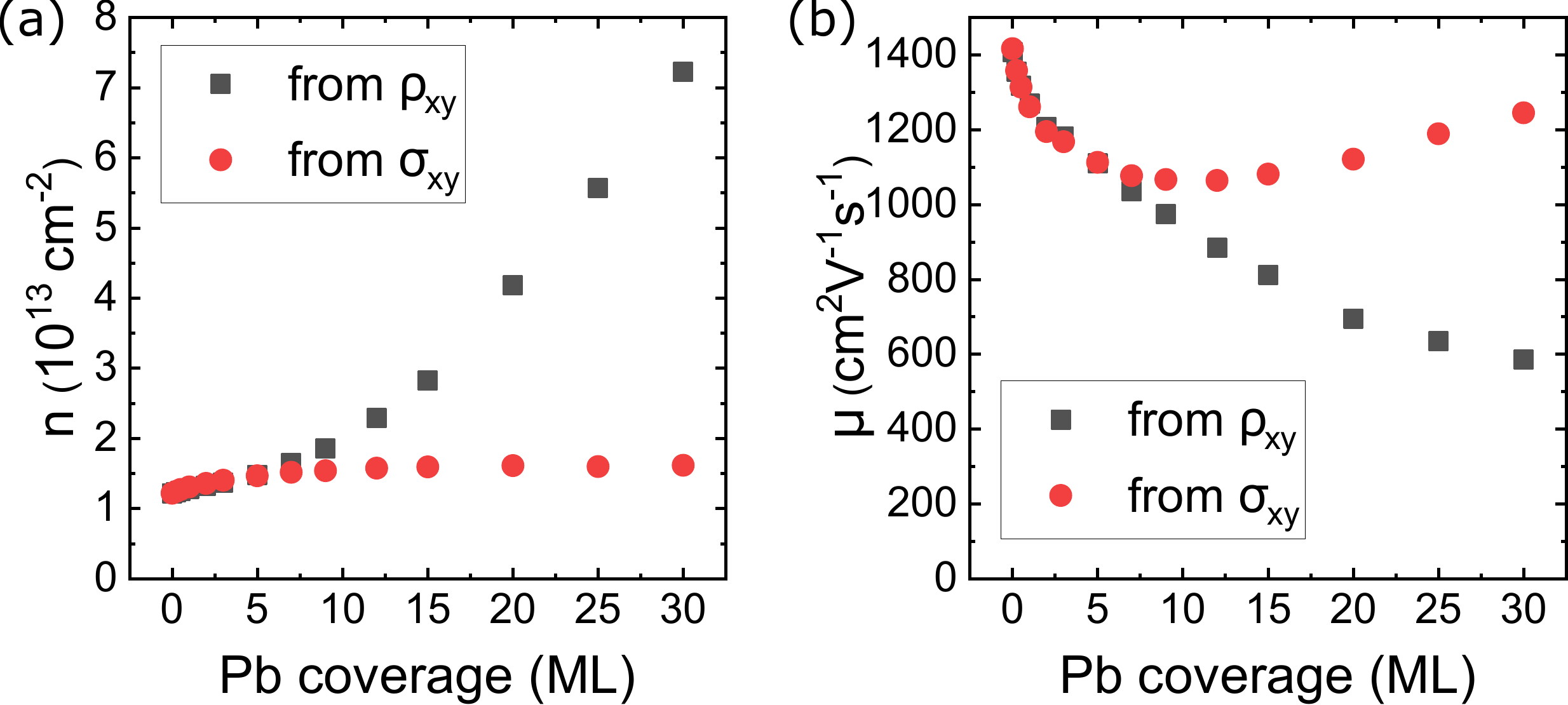}
		\caption{\label{fig:Magneto_singleband} 
		Single-band model: (a) Electron concentration and (b) mobility in dependence of the average Pb coverage extracted from $\rho_{xy}$ (black squares) and $\sigma_{xy}$ (red circles) using the single-band model (see text for details).
		%G1745
		}
	\end{center}
\end{figure}

\subsubsection{Classical contribution}

The magnetoresistivity for average Pb coverages of 0 to 30\;ML is shown in Fig.~\ref{fig:Magneto_overview}(a). At low magnetic fields $\abs{B}<1\;\text{T}$ a WL peak is clearly visible, whose intensity decreases with increasing Pb coverage. At magnetic fields $\abs{B}>2\;\text{T}$ the shape of the magnetoresistivity curves remains similar up to 3\;ML, except for small changes at the lowest coverages. Starting at 5\;ML, the quasi-parabolic contribution starts to increase with increasing coverage. A quantum correction to the magnetoresistivity with a quasi-parabolic magnetic field dependence can arise in graphene due to electron-electron interaction (EEI) (see, e.g., Refs.~\onlinecite{Jouault2011,Jobst2012,Kozikov2010}). This is the origin of the quasi-parabolic background already present for MLG. However, the quasi-parabolic contribution that dominates the magnetoresistivity at higher coverges cannot be explained by EEI, since it does not show a strong temperature dependence, as demonstrated in Fig.~\ref{fig:Magneto_T-dependence} for a coverage of 12\;ML. Consequently, the increase of the quasi-parabolic contribution for $\geq5\;\text{ML}$ has to be of classical origin. The Hall resistivity, shown in Fig.~\ref{fig:Magneto_overview}(b), remains linear for all measured coverages, with its absolute slope decreasing with increasing coverage. This decrease indicates an increase of the carrier concentration, which will be discussed in detail further below. 

Neglecting any quantum corrections, in particular WL and EEI, a homogeneous two-dimensional electron gas with only one type of charge carrier, such as graphene, shows no magnetoresistance, i.e., $\rho_{xx}(B)=const.$~\cite{Pippard1989}. Therefore, the classical quasi-parabolic contribution at $\geq 5\;\text{ML}$ means that the system cannot be described as such in this coverage range. In order to confirm this, we determined the electron concentration $n$ and the mobility $\mu$ from both $\rho_{xy}$ and $\sigma_{xy}$ assuming homogeneity and only one type of charge carrier. In the first case, $n$ was determined from the slope of $\rho_{xy}$ via $\rho_{xy}(B)=-B/(ne)$, with $e$ being the elementary charge, and $\mu$ was subsequently determined from the conductivity at zero magnetic field $\sigma_0=ne\mu$. In the second case, both $n$ and $\mu$ were determined from a fit of $\sigma_{xy}$ using $\sigma_{xy}(B)=ne\mu^2B/(1+\mu^2B^2)$. If this single-band model is sufficient to describe the data (the WL contribution can be ignored here, since it does not contribute to $\rho_{xy}$ and its contribution to $\sigma_{xy}$ is negligible), the results from both methods have to agree with each other. As expected, this is the case for coverages below 5\;ML, while at higher coverages there is an increasing disagreement between the results from the two methods, as seen in Fig.~\ref{fig:Magneto_singleband}. 

The classical quasi-parabolic contribution to the magnetoresistivity for coverages $\geq 5\;\text{ML}$ is indicative of either the emergence of a second charge carrier type~\cite{Pippard1989}, i.e., transport through the Pb islands, or scattering of the charge carriers due to an inhomogeneous carrier concentration or mobility~\cite{Shik2004,Polyakov2001}. We previously observed a similar contribution for Bi islands on MLG/SiC(0001), which we ascribed to an inhomogeneous carrier concentration~\cite{Koch2024}. However, the key difference to the present case is that in the case of Bi this contribution exists already at the lowest coverages, where there is no percolation of the adsorbed Bi. In contrast, in the case of Pb islands on MLG/SiC(0001) the classical quasi-parabolic contribution arises between 3 and 5\;ML, where percolation can reasonably be expected to occur based on the minimum in $\sigma_0$ (see bottom left inset in Fig.~\ref{fig:Magneto_overview}(b)) and the STM investigation, which showed a transition from the growth of small, similarly sized islands to larger islands with a greater variance in size in this region. In addition, in the case of the Pb islands the onset of this contribution is correlated with the scattering lengths becoming independent of the Pb coverage (see Fig.~\ref{fig:Magneto_scattering} and the discussion of the WL below). In contrast, in the case of Bi islands the increase of the classical quasi-parabolic contribution is correlated with a strong decrease of the coherence time, due to scattering at the island edges. For these reasons, we ascribe the increase of the quasi-parabolic contribution for $\geq5\;\text{ML}$ in the case of the Pb films to transport through the Pb layer.

\begin{figure}[tb]
	\begin{center}
		\includegraphics[width=1\columnwidth]{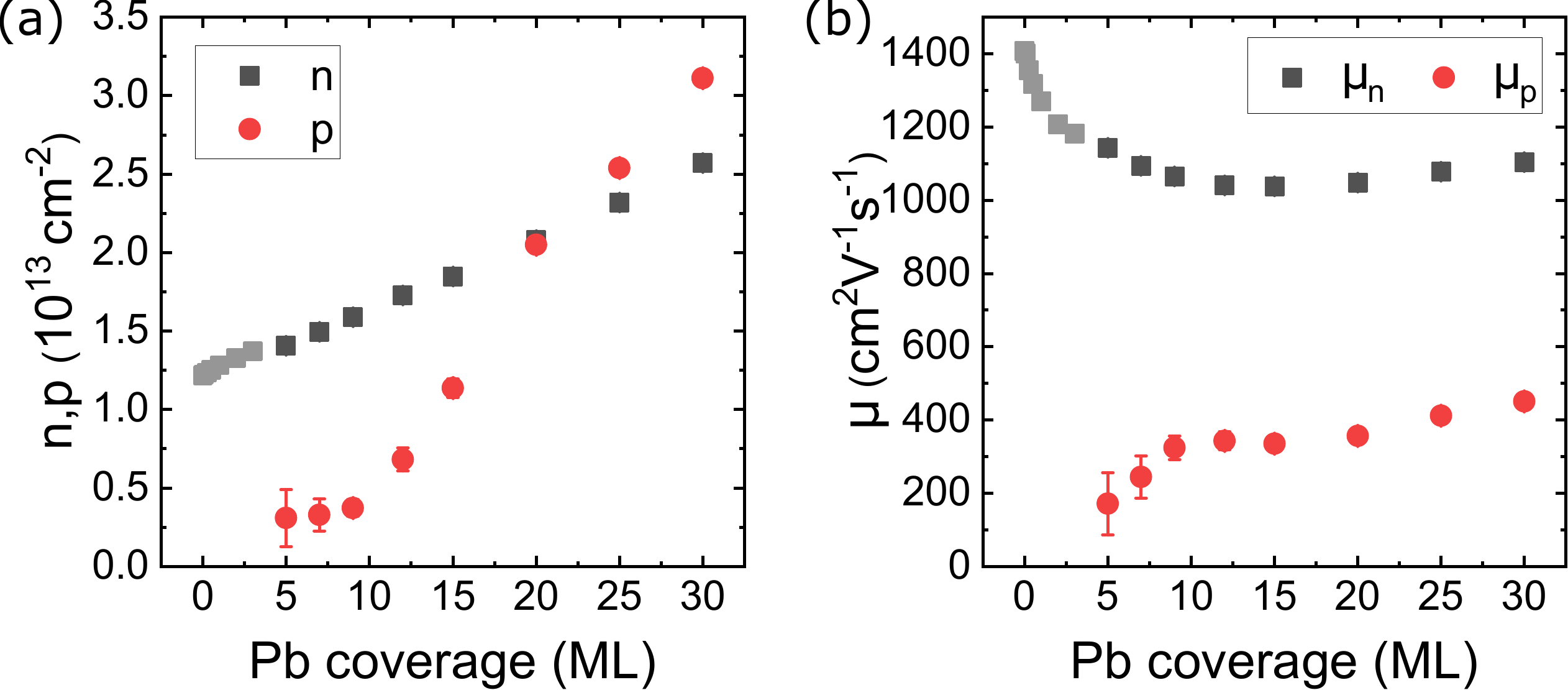}
		\caption{\label{fig:Magneto_twoband} 
		Two-band model: (a) Electron concentration $n$ and hole concentration $p$, as well as (b) the respective mobilities $\mu_n$ and $\mu_p$ in dependence of the average Pb coverage. The values for coverages $\leq 3\;\text{ML}$ (light gray symbols) were obtained using the single-band model. 
		%G1745
		}
	\end{center}
\end{figure}

Furthermore, a consistent description of the data with a two-band model, in which it is assumed that both electrons and holes contribute to the overall carrier transport, is possible in the case of Pb islands, but did not give a reasonable fit in the case of Bi islands. In the two-band model the magnetic field dependence of the magnetoresistivity $\rho_{xx}$ and the Hall resistivity $\rho_{xy}$ are given by~\cite{Pippard1989}
\[
\rho_{xx}(B)=\frac{1}{ne}\frac{\mu_n+c\mu_p+\mu_n\mu_p(\mu_p+c\mu_n)B^2}{(\mu_n+c\mu_p)^2+(1-c)^2\mu_n^2\mu_p^2B^2}
\]
and 
\[
\rho_{xy}(B)=-\frac{B}{ne}\frac{\mu_n^2-c\mu_p^2+(1-c)\mu_n^2\mu_p^2B^2}{(\mu_n+c\mu_p)^2+(1-c)^2\mu_n^2\mu_p^2B^2},
\]
respectively, where $n$ is the electron concentration, $p$ is the hole concentration, $c=p/n$, and $\mu_n$ and $\mu_p$ are the respective mobilities. A consistent description of the data in this context means that both $\rho_{xx}$ and $\rho_{xy}$ are described by a single set of parameters $n, p, \mu_n, \mu_p$ that simultaneously satisfies the condition $\sigma_0=ne\mu_n+pe\mu_p$, where $\sigma_0$ is the conductivity at zero magnetic field. 

The results of the fitting process are shown in Fig.~\ref{fig:Magneto_twoband}. The values for coverages $\leq 3\;\text{ML}$ were obtained using the single-band model. The electron concentration $n$ increases linearly with approximately $4.8\times10^{11}\;\text{ML}^{-1}\text{cm}^{-2}$ across the entire coverage range. The hole concentration $p$ increases with $1.4\times10^{12}\;\text{ML}^{-1}\text{cm}^{-2}$ starting at 9\;ML, becoming larger than $n$ at around 20\;ML. The electron mobility $\mu_n$ sharply decreases below 5\;ML from its inital value of $1400\;\text{cm}^2\text{V}^{-1}\text{s}^{-1}$ for the bare MLG, and then saturates at around $1050\;\text{cm}^2\text{V}^{-1}\text{s}^{-1}$. The hole mobility $\mu_p$ is generally lower than $\mu_n$, saturating at around $400\;\text{cm}^2\text{V}^{-1}\text{s}^{-1}$. The hole mobility is in the same order of magnitude as the mobility in thin Pb film on Si(111)~\cite{Luekermann2013}, and much lower than the typical mobility in graphene. Moreover, additional hole contributions are unlikely to arise in graphene at the doping level of our substrates, since there are no bands close to the Fermi level except at the $K/K^\prime$ points. Therefore, we ascribe the hole contribution to the Pb states. The initial increase of the hole mobility at low coverages is an indication that the percolated path has not yet fully formed in this coverage range, so that the cross section at the connection points of the islands still limits the transport.

\begin{figure}[tb]
	\begin{center}
		\includegraphics[width=1\columnwidth]{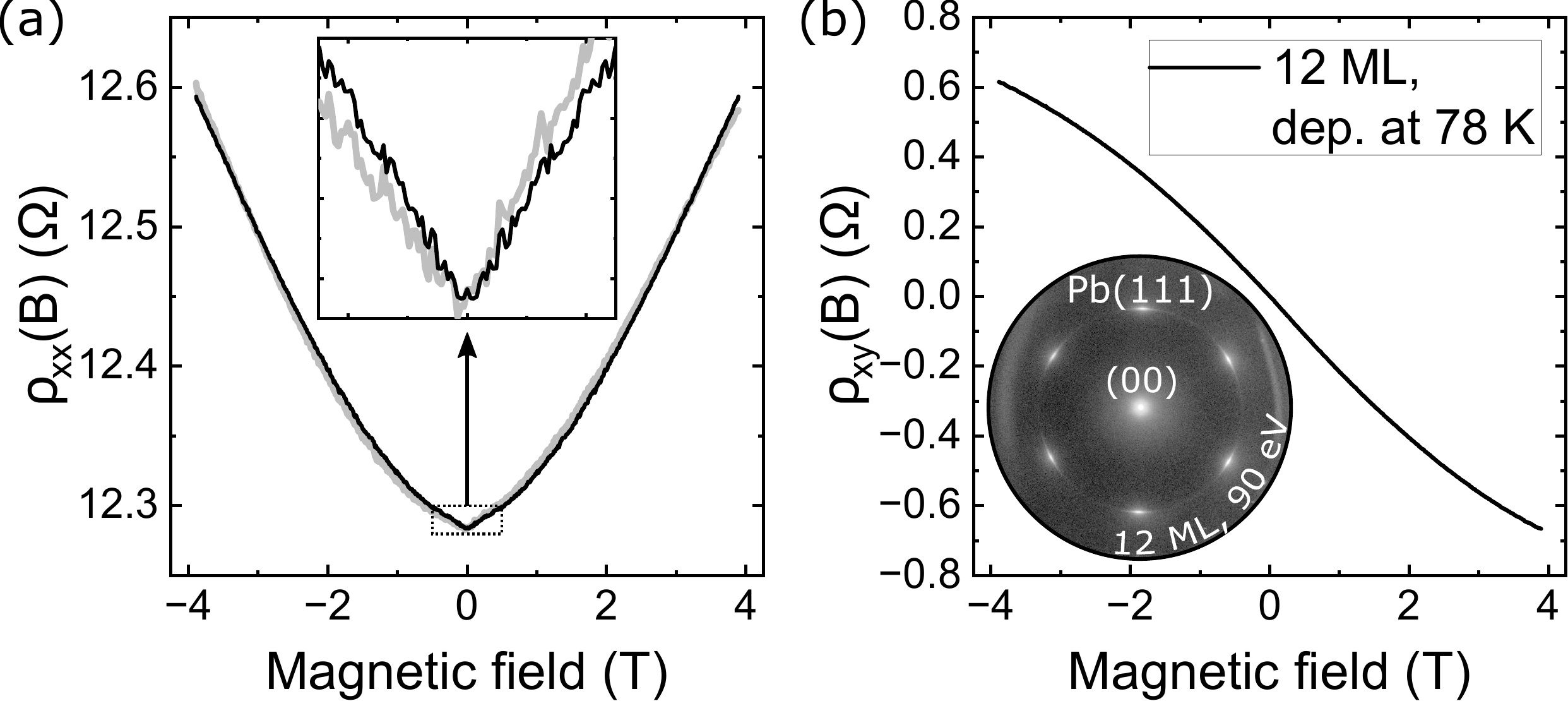}
		\caption{\label{fig:lN2-deposition} 
		(a) Magnetoresistivity and (b) Hall resistivity measured at 12\;K of a 12\;ML thick Pb(111) film deposited on MLG/SiC(0001) at a deposition temperature of 78\;K. The black curve in (a) was symmetrized with respect to $B=0$, i.e., $(\rho_{xx}(B)+\rho_{xx}(-B))/2$. The light gray curve shows the original data. The inset in (a) shows a magnification of the WAL behavior at low magnetic fields. The inset in (b) shows the LEED image of the same film recorded at 78\;K using an electron energy of 90\;eV.  
		%The distortion of the LEED image was corrected using LEEDCal~\cite{LEEDCal,Sojka2013}.
		%G1745
		}
	\end{center}
\end{figure}

The Hall resistivity appears linear for all Pb coverages investigated, even though the data is consistent with the two-band model. This requires the conditions $\mu_n^2-c\mu_p^2 \gg (1-c)\mu_n^2\mu_p^2B^2$ and $(\mu_n+c\mu_p)^2 \gg (1-c)^2\mu_n^2\mu_p^2B^2$ to be fulfilled across the entire range of coverages, which is indeed the case for the values shown in Fig.~\ref{fig:Magneto_twoband}. 
It should be noted at this point, that the second type of charge carriers that arises has to be hole-type. Assuming a second type of electrons with a mobility that differs from the one of the graphene electrons instead, would reverse the sign of the respective mobility in the above equations (cf.\ Ref.~\onlinecite{Pippard1989}). The model then no longer gives a consistent fit of the data. 
Furthermore, possible contributions from electron states in the Pb(111) islands were neglected in the analysis, because their consideration would require a three-band model, thus raising the number of fitting parameters to such an extent that a unique fit is no longer possible. Moreover, they appear to be unnecessary for a description of the data. The Fermi surface of bulk Pb has equal electron and hole concentrations~\cite{Alekseevskii1962}. However, in thin films either electrons or holes dominate depending on the layer thickness due to the quantum size effect~\cite{Vilfan2002}. In particular, according to Ref.~\onlinecite{Vilfan2002} holes dominate the transport between 7 and 10\;ML (higher thicknesses were not investigated), which is consistent with our results. 

For comparison, we also measured a closed Pb(111) film on MLG/SiC(0001) with a thickness of 12\;ML (see Fig.~\ref{fig:lN2-deposition}). As it turned out, it is possible to grow a closed film at a sample temperature of 78\;K. The magnetoresistivity and Hall resistivity were measured after cooling the sample with liquid helium directly after the growth without letting the sample heat up, since the closed film was so unstable that even heating it to room temperature would result in the film tearing up. 
The Pb(111) film/MLG stack shows a conductivity of $\sigma_0=81.4\;\text{mS}$, which is an increase by more than one order of magnitude as compared to the case of percolated islands, indicating that the transport is now dominated by the Pb layer. The Hall resistivity now shows the typical S-shape, expected for the contribution of both electrons and holes to the transport. Moreover, the magnetoresistivity of the closed film shows WAL behavior as seen in the inset of Fig.~\ref{fig:lN2-deposition}(a), which was also observed for Pb(111) films on Si(111)~\cite{Pfennigstorf2002}.

\subsubsection{Weak localization}

A comprehensive model of the WL/WAL quantum correction of the magnetoresistivity in graphene, which also considers SOC, is given by McCann and Fal’ko in Ref.~\onlinecite{McCann2014}. However, due to the consideration of five scattering rates, associated with the coherence of the electrons, intervalley scattering, intravalley scattering, Kane-Mele SOC and Bychkov-Rashba SOC, respectively, simplifying assumptions have to be made in order to reduce the number of fitting parameters. 
Firstly, the influence of the Kane-Mele type SOC on the WL is negligible for our samples, since even for the largest SOC gaps of around 21\;meV, that can be realistically expected~\cite{Weeks2011}, the Kane-Mele scattering rate is at least two orders of magnitude smaller than typical values of the coherence scattering rate in our substrates~\cite{Koch2024}. 
Secondly, according to the model, the Bychkov-Rashba scattering rate becoming non-negligible compared to the coherence rate would result in an abrupt switch from WL to WAL (assuming that the inter-\ and intravalley scattering rates do not change drastically). However, a gradual suppression of WL was observed. Therefore we can assume the influence of the Bychkov-Rashba SOC to be negligible as well. 
We confirmed this by fitting our data with the simplified model from Ref.~\onlinecite{McCann2012}, which groups all scattering rates in those originating from either a broken (asymmetric) or non-broken (symmetric) mirror symmetry with respect to the graphene plane. The result was that the asymmetric term, and therefore the Bychkov-Rashba SOC, is negligible for the entire coverage range. 

The absence of Bychkov-Rashba SOC seems to contradict the theoretical expectation. A combination of analytical calculations and tight-binding simulations predicts that adatoms on graphene in the top position, which is the preferred adsorption site of individual Pa adatoms~\cite{Hupalo2011}, as well as random distributions of Pb adatoms induce Bychkov-Rashba SOC, whereas only adatoms exclusively in hollow positions do not induce Bychkov-Rashba SOC but increase the intrinsic Kane-Mele SOC~\cite{Brey2015}. 
The distance between graphene hollow positions is 2.46\;{\AA}, i.e., the graphene lattice constant, in zigzag direction and 4.27\;{\AA} in armchair direction. Both distances are incommensurate with the Pb(111) lattice constant of 3.5\;{\AA}, so that all Pb atoms being in hollow positions is not expected for Pb(111) on graphene. Consequently, the Pb(111) islands should induce Bychkov-Rashba SOC. 
However, for Pb deposited on graphene/Ru(0001) a $(2\times2)_{Gr}$ superstructure with all Pb atoms being in hollow positions was found for single layer Pb islands~\cite{Liu2013}. If this structure is maintained at the interface between the Pb(111) islands and graphene, an absence of Bychkov-Rashba SOC can be rationalized. On the other hand, a model of individual adatoms might not be suitable to predict the SOC induced by Pb islands with an extended band structure, which allows for a rearrangement of the charge inside the island that can affect the dominant tunneling channels.

\begin{figure}[tb]
	\begin{center}
		\includegraphics[width=1\columnwidth]{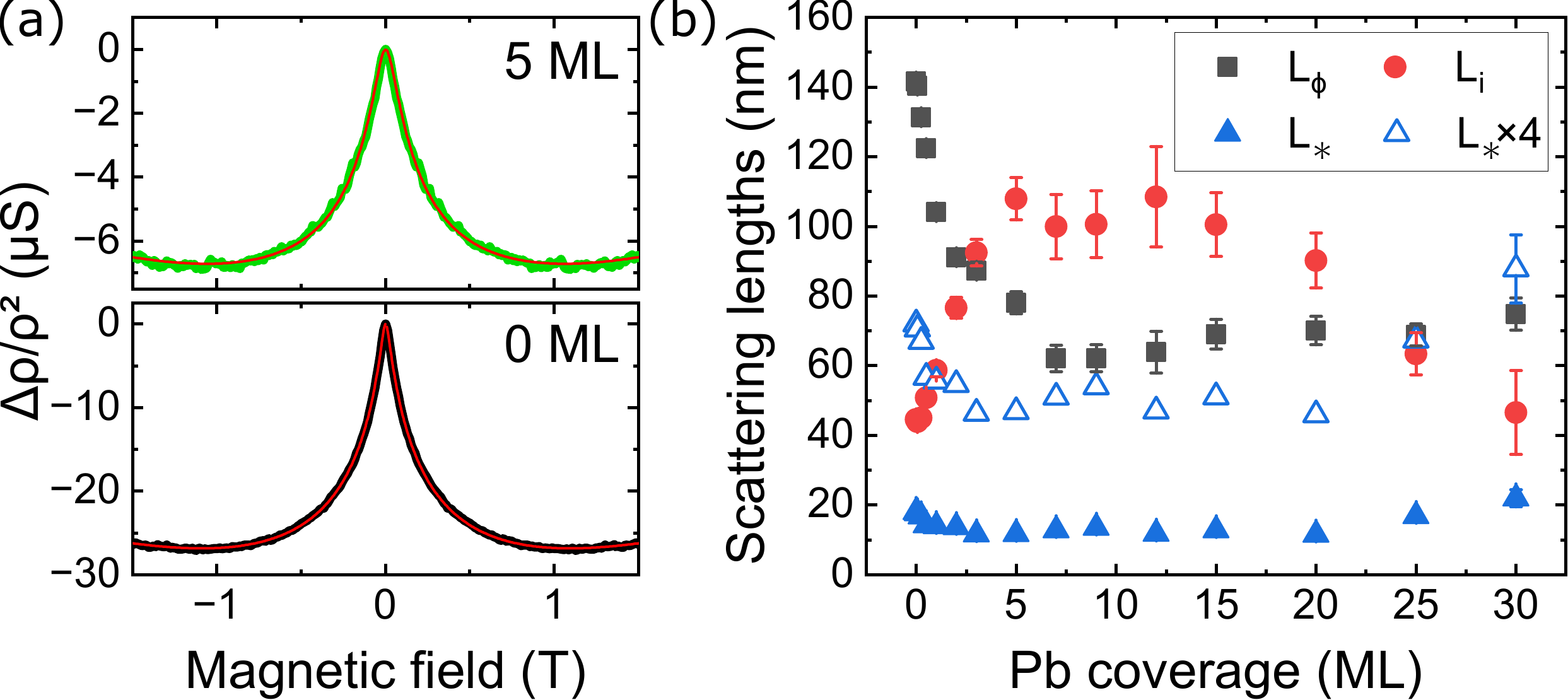}
		\caption{\label{fig:Magneto_scattering} 
		(a) Fits of the WL (red) for the clean graphene surface (black) and for a Pb coverage of 5\;ML (green). See text for details. (b) Coherence length $L_\varphi$, intervalley scattering length $L_i$ and intravalley scattering lenght $L_*$ in dependence of the average Pb coverage. For better visibility, the intravalley scattering lenght is also shown multiplied by a factor of four using unfilled symbols.
		%G1745
		}
	\end{center}
\end{figure}

Since both Kane-Mele SOC and Bychkov-Rashba SOC were found to be negligible, we also analyzed our data with the WL/WAL model, that does not consider SOC. The WL/WAL quantum correction to the magnetoresistivity in graphene without considering SOC is given by~\cite{McCann2014,Kechedzhi2007}
{
\medmuskip=-1mu
\thinmuskip=0mu
\thickmuskip=0mu
\nulldelimiterspace=0pt
\scriptspace=0pt
\begin{multline*}
\frac{\Delta \rho(B)}{\rho^2}=-\frac{e^2}{\pi h}\left[F\left(\frac{B}{B_\varphi}\right)-2F\left(\frac{B}{B_\varphi+B_*}\right)-F\left(\frac{B}{B_\varphi+2B_i}\right)\right],
\end{multline*}
}
with
\begin{align*}
&B_x=\frac{\hbar c}{4e}L_x^{-2}, \quad x=\varphi, i, *,
\end{align*}
where $F$ and $L_\varphi$ are defined as above, $L_*^{-2}=L_w^{-2}+L_v^{-2}+L_i^{-2}$, $L_i$ is the intervalley scattering length, $L_v$ is the intravalley scattering length and $L_w$ is the trigonal warping relaxation length. Due to $L_v$ being the dominant contribution to $L_*$, $L_*$ is usually considered equivalent to the intravalley scattering length. 
Fits for the clean graphene surface and a Pb coverage of 5\;ML are shown in Fig.~\ref{fig:Magneto_scattering}(a) as examples. A parabolic background was used to compensate for the EEI contribution. For coverages $\geq5\;\text{ML}$ the classical contribution was also taken into account. 

The scattering times resulting from the fitting process are shown in Fig.~\ref{fig:Magneto_scattering}(b). Below 5\;ML, i.e., below the percolation threshold for Pb, both the coherence length and the intravalley scattering length decrease as expected. In contrast, the intervalley scattering length actually increases, i.e., the intervalley scattering rate decreases. Intervalley scattering in graphene is mediated by atomic defects and therefore absent in ideal graphene~\cite{Tikhonenko2009}. Of course the defect density does not decrease due to the adsorption of Pb, as is evident from the decrease of the mobility. Therefore, the decrease in intervalley scattering has to be the result of local changes in the scattering potential. Indeed, Zhang \textit{et al.}\  have shown that the charging of monovacancies in graphene can locally soften the scattering potential and thereby suppress intervally scattering~\cite{Zhang2022}. It is reasonable to expect the Pb to cause such a charging, given the observed overall charge transfer to graphene and the expectation that the island growth starts primarily at defects due to the inertness of graphene. 

At larger coverages, where there is percolation of the adsorbed Pb, the scattering lengths saturate and eventually the initial changes seem to be reversed. In principle, an additional WL/WAL contribution originating from the Pb can be expected in this coverage range. Pb(111) shows WAL behavior as we have shown above (see Fig.~\ref{fig:lN2-deposition}), which if present in addition to the WL in the graphene would result in an underestimation of the WL peak, and therefore the coherence length, originating from graphene. However, we observe a saturation of the coherence length instead of a continued decrease. Moreover, the magnitude of the WAL observed in the closed Pb(111) film is more than one order of magnitude smaller than that of the WL observed in the graphene covered by Pb islands at even the highest coverages. Therefore, the influence of the WAL contribution from Pb states is most likely negligible. The evolution of the scattering lengths in this regime is consistent with additional Pb being deposited predominantly on top of existing islands, where it cannot affect the scattering behavior in the graphene, and the island edges, whose density decreases due to the formation of larger islands and the adsorbed Pb approaching full coverage, acting as scattering centers.

%_CONCLUSION__________________________________________________________________________________________________________________________________________
\section{Summary and Conclusion}

In conclusion, Pb(111) islands act as electron donors increasing the electron concentration of the graphene by about $5\times10^{11}\;\text{ML}^{-1}\text{cm}^{-2}$. These electrons can move freely in the whole graphene layer and there is no evidence that the adsorbed Pb islands induce locally varying electron concentrations. This is in stark contrast to the case of Bi(110) islands on graphene, which only increase the electron concentration in the graphene directly below the islands and leave the electron concentration in the remaining graphene layer unaffected~\cite{Koch2024}. 
Above the percolation threshold, hole transport through the Pb layer has to be taken into account in order to describe the data. 

The Pb islands do not induce any significant Rashba-type spin-orbit coupling. This result is only consistent with theoretical calculations, if it is assumed that all Pb atoms at the interface are in hollow positions of the graphene lattice~\cite{Brey2015}. A $(2\times2)_{Gr}$ Pb superstructure, that fulfills this condition, was indeed observed on graphene/Ru(0001) for single layer islands~\cite{Liu2013}. Our results suggest that this or a comparable structure is maintained at the interface between Pb(111) and graphene. 
Furthermore, the adsorbed Pb seems to screen the defects in the graphene, thus softening the scattering potential and decreasing the intervalley scattering rate in the graphene layer. 

This work demonstrates that an inhomogeneous coverage on graphene does not necessarily result in an inhomogeneous electron concentration, and underlines the importance of the resulting doping profile in proximitized graphene for a successful functionalization. 
An inhomogeneous or homogeneous doping profile might also be the reason for the apparent increase in line width in photoemission spectroscopy observed after the decoration of graphene with some elements but not others~\cite{Gierz2008}.

%################################################################################################################

\vspace{1ex}
{\bf Acknowledgement} We thank C. Lohse (TU Chemnitz) for providing us with the graphene sample for the STM measurements. This work was supported by the Deutsche Forschungsgemeinschaft (DFG) projects Te386/22-1 and Pi385/3-1 within the FOR5242 research unit.\\

%\bibliography{magneto}


\end{document}